 
   \magnification=\magstep1

   \font\stephalf=  cmr6 scaled\magstep0  
 \font\abstract=  cmr7 scaled\magstep1

\settabs 18 \columns
 \hsize=16truecm
  
\input epsf

\def\s{\sigma}

\def\b{\bigskip}
\def\bb{\bigskip\bigskip}

\def\no{\noindent}
\def\r{\rightline}
\def\ce{\centerline}
\def\ve{\vfill\eject}

\def\r{\rightline}

\def\s{\sigma}

\def\harr#1#2{\smash{\mathop{\hbox to .25 in{\rightarrowfill}}
 \limits^{\scriptstyle#1}_{\scriptstyle#2}}}

\def\today{\ifcase\month\or January\or February\or March\or
 April\or
May\or June\or July\or
August\or September\or October\or
 November\or  December\fi
\space\number\day, \number\year }

\r \today

\def\e{\eta}

\def\p{\partial}

\def\sqr#1#2{{\vcenter{\vbox{\hrule height.#2pt
\hbox{\vrule width.#2pt height#2pt \kern#2pt
\vrule width.#2pt}
\hrule height.#2pt}}}}

\def\picture #1 by #2 (#3){
  \vbox to #2{
    \hrule width #1 height 0pt depth 0pt
    \vfill
    \special{picture #3} 
    }
  }
 %
%

\font\steptwo=cmb10 scaled\magstep2
\font\stepthree=cmb10 scaled\magstep4
\magnification=\magstep1

  \ce{\stepthree   Stability of Polytropes}\b
  \ce{Christian Fr\o nsdal}
\b
  \ce{\it Physics Department, University of California, Los Angeles CA
  90095-1547 USA}
\vskip1.5cm

\def\sqr#1#2{{\vcenter{\vbox{\hrule height.#2pt
\hbox{\vrule width.#2pt height#2pt \kern#2pt
\vrule width.#2pt}
\hrule height.#2pt}}}}

\def \r{\rightarrow}

\def\e{{\rm e}}
 \no{\it ABSTRACT.} This paper is an investigation of the stability of some
ideal stars.    It is intended as a study in General Relativity, with
emphasis on the coupling to matter, eventually aimed at a better
understanding of very strong gravitational fields and ``Black Holes".
This contrasts with the usual attitude in astrophysics, where Einstein's
equations are invoked as a refinement of classical thermodynamics and
newtonian gravity. Our work is based on an action principle for the
complete system of metric and matter fields, coupling the metric field
to well defined relativistic field models that we hope may  represent a
plausible type of matter. The thermodynamic content must be extracted from
the theory itself.   We start with the simplest model of matter and plan
to add complications as our experience grows.  When the flow  of matter is
irrotational, and described by a scalar density, we are led  to
differential equations that differ little from those of Tolman and his
followers, but equations derived from an action principle admit a
conserved current, as well as stronger boundary conditions that affect the
matching  of the
interior solution to an external Schwarzschild metric  and imply a
relation between mass and radius.

  We have been led to propose a complete revision of the
treatment of boundary conditions.   An ideal star in our terminology has
spherical symmetry  and an isentropic equation of state, $ p
=a\rho^\gamma$, $a$  and $\gamma$ piecewise constant.
In our first work on this subject it was assumed that the density
vanishes beyond a finite distance from the origin and that the metric
is to be matched continuously at the boundary to an exterior
Schwarzschild metric. But it is difficult to decide what the proper
boundary conditions should be and we are consequently skeptical of
the concept of a fixed boundary. In this paper we
investigate
the double polytrope, characterized by a polytropic index  $n \leq 3$
in  the
bulk of the star and a value larger than 5 in an outer atmosphere that
extends to infinity. It has no fixed
boundary but a region of critical density where the polytropic index
changes from a value that is appropriate for the bulk of the star to a
value that provides a crude model for the atmosphere,
   with a fairly abrupt transition at the point where the density
reaches a critical value. All the fields are continuous. The boundary 
conditions are now natural and
unambiguous. The existence
of a
relation between mass and radius is confirmed, as well as an upper limit
on the mass.  The principal  conclusion is
that all the static configurations are stable. There is a solution that
fits the sun. The masses of white dwarfs respect the Chandrasekhar limit.
The application to neutron stars has suprising aspects.

\ve

\no{\steptwo 1. Introduction }

The problem on which we hope to throw some light in this paper is the
application of General Relativity to star-like systems that can be
described by mass, radius, density and pressure; the flow is assumed to
be irrotational and to be controlled by a velocity potential.

In contrast with the traditional treatment we introduce the matter
component into Einstein's equation for the metric by adding an
appropriate matter contribution to the\break  Einstein-Hilbert action.
The difference, at first sight, seems minor, for the equations
associated with either approach are nearly identical, especially in the
static case. An important consequence for the study of equilibrium
configurations is that the action principle fixes an integration
constant
that is left free in the phenomenological approach.  This
results in a strengthening of the conditions for matching the
interior metric to an exterior Schwarzschild metric at the boundary of
the star - see Eq.(1.4), or  asymptotically at
infinity.
Consequently, the action principle has some additional predictive
power, likely to bring it down, perhaps, but worth investigating.

We do not take it for granted
that the matter distribution conforms to the precepts of classical
thermodynamics; instead it is expected that the interpretation is
implied
by the theory itself. It turns out that a simple choice of
interaction does indeed lead to an equation of state of the familiar
type.   The
difference in attitude therefore does not, by itself, lead to any
dramatic  contrast with the
semi-phenomenological approach. The future
inclusion of
radiative effects may change that.

  The simplest
choice of action leads to  an equation of state of the form
$$
p = {a\over n} \rho ^\gamma,~~ \gamma = 1 + {1\over n},\eqno(1.1)
$$
where    $a$,$\gamma$ are constants,
eventually piecewise constant.

\b

The main  result of an earlier investigation [F2] was a relation
between the mass and the radius of any equilibrium configuration. The
original purpose of the present paper was to study the stability of
those
configurations. In the course of this work we have become somewhat
skeptical about the appropriateness of naive  boundary
conditions; that is,  the assumed continuity of the metric, regardless
of the behaviour of density and pressure that it implies, at a fixed
boundary. A large part of this paper is directed to a re-evaluation  of
these questions.
   \b

\ce {\bf The double polytrope}

Consider the process that leads to the formation of a star, assuming
that
the primordial matter is homogeneous, as is reasonable if
stars are a late result of a long process of
condensation of a hydrogen cloud. Condensation is a result of
gravitational attraction and the first effect produced by the
attraction is
an increase in density. All subsequent development is ultimately
attributable
to this primordial increase in the density. If, as is always taken for
granted in studies of stellar structure, the equation of state is
nearly
isentropic, then the basic, underlying  reason for a change in the
index
$n$ must be the variation of density. A popular model is a polytrope with
$n\approx 3$ in a region of moderately high density and $ n>5$ in
the outer atmosphere. In the early stages with low density the
index may be larger than 5 almost everywhere, characteristic of a
distribution that extends to infinity, but as this would imply a
singularity at the origin a change must take place near the center.
  Whatever happens,  the primal cause is
the variation of density. That is, the position of the boundary must be
determined by the density, rather than the other way around. It follows
that, if the index changes abruptly, then it is the result of a rapidly
changing density, as in
$$
n[\rho] = {n_1(\rho /\rho_{cr})^K + n_2\over 1 +(\rho
/\rho_{cr})^K},\eqno(1.2)
$$
where $K$ is a suitable large number and $\rho_{cr}$ is a critical density.
(Another representation for the approximately piecewise function will be
explored at the end, with interesting results.)

Indeed, in an approximation where the only variables to be taken into
account, besides the components of the metric field, are density and
pressure, this would appear to be the only possible approach: {\it the
boundary is defined to be the region of critical density}. Note that this
``boundary" need not coincide with the visual boundary of the star.

In this paper, after attempting a more traditional approach to
localizing the surface of the star, and remaining unconvinced of the
aptness of it, we shall concentrate on trying to understand the double
polytrope with this type of equation of state.

The main conclusion is that all the static solutions, with natural
boundary conditions applied at the center and at infinity, are stable.
The white dwarfs
respect the Chandrasekhar limit on the mass, not because heavier stars are
unstable, but because they do not exist.
A model for the sun is included without any surprises; the
application to neutron stars offers new dimensions.

\b

\ce{\bf Summary }

Section 2. An unfamiliar aspect of this work is the use of an
action principle for the  complete system of metric and matter fields.
Matter is assumed to be irrotational and isentropic, thus fully described
by fields of density and pressure, with the action
$$
A_{matter} = \int d^4x\sqrt{-g}\Big({\rho\over 2}( g^{\mu\nu} \psi_{,\mu}
\psi_{,\nu} - c^2) - V[\rho]\Big) =:  \int d^4x\sqrt{-g}~{\cal
L}~.\eqno(1.3)
$$
The eventual presence of (electromagnetic) radiation will require
additional terms to be added. The associated energy momentum tensor
provides the right hand side of Einstein's equation $R_{\mu\nu} -
g_{\mu\nu} R = 8\pi GT_{\mu\nu}$.   We look for solutions that are
spherically symmetric,   such that the metric, in terms of coordinates
$t,r,\theta$ and $\phi$,   takes the form
   $$
(ds)^2 = {\e}^\nu(dt)^2 -  \e^\lambda(dr)^2  - r^2d\Omega^2,
$$
with $\nu$ and $\lambda$ depending on $r$ and $t$ only.

Homogeneous polytropes are characterized by a potential (the internal
energy) $V[\rho]$ of the form $V[\rho] = a\rho^\gamma$, with $\gamma$
and $a$ constant,  which leads to the equation of state $p =
a(\gamma-1)\rho^\gamma$. When $\gamma$ is not constant the polytropic
equation of state is slightly modified in the region of critical
density. As is usual, we study the time development of  the system under
the assumption that it is initially in an equilibrium configuration.

Our first calculations [F2] postulated a fixed boundary
beyond which the density is zero and the metric is that of
Schwarzschild's exterior solution. The static configurations of this model
are essentially the same as in the traditional, phenomenological
approach, except for an important  difference with respect to the
boundary conditions. In contrast with the situation in the usual
approach, we must match both of the metric functions $\nu,\lambda$
of the interior solutions  
to an external Schwarzschild metric. Integration proceeds from the
center and the boundary is at a point $r = R$ where
$$
\nu(R) + \lambda(R) = 0.\eqno(1.4)
$$
Within the traditional approach this condition is ineffective since
the boundary value of $\nu$ is just an integration constant.

  We wish to calculate the time development to first order in the
deviations from equilibrium. This was first done by Chandrasekhar in
1931 [C1], and as far as we know the same method has been followed by
all
later investigators. All these studies are characterized by what we
think are insufficiently motivated boundary conditions. In the first
place it is  not
sure that one knows which of the fields, metric components, density,
pressure, should  be required to be continuous at the boundary.
The boundary is not at a fixed  point
but varies from one static configuration to another and with time.
Consequently it is unnatural to restrict the fluctuations by the
condition that the radius remain fixed. Another question that imposes
itself is that of the  mass. We define the mass in terms of the
asymptotic metric; does it echo the oscillations or does it remain fixed?

In view of the fact that we have come to view the Schwarzschild
solution
as the metric of a singular, limiting  mass distribution [F1][F2][MM], we
felt that one way to clarify these questions would be to replace the outer
Schwarzschild metric with another polytrope, with index $n > 5$ as is
appropriate for a mass distribution that extends to infinity. This
exterior metric rapidly approaches the Schwarzschild metric at moderate
distances. But the boundary conditions continue to present the same,
difficult  problems.

Section 3. We have  come to believe that the very idea of a fixed boundary
is unnatural, and an obstruction to understanding what is going on.
For this
reason we make a new start, with another version of the double polytrope,
an ideal star in which the equation of state is essentially  polytropic
near the center, with an index that is nearly constant, but changes
to a larger value in a ``boundary" region of critical density, and
essentially  constant outside this region.
   The ``surface"  of this star is a  region in which the index
makes a sudden or gradual change from one value to the other, as in
$$
p ={a\over n}f^{n+1},~~\rho = f^n,~~n = 6- 3 {(f /f_{cr})^K\over 1
+(f /f_{cr})^K}.\eqno(1.5)
$$
This defines a 2-parameter family of equations of state. For reasons that 
will be explained, the most plausible models have $f_{cr} = \rho_{cr} =
1$.

\ve

\b\ce{\bf Results}

   Our earlier calculations were made with a fixed isentropic
index for the interior, and a matching Schwarzschild metric for the empty
exterior. In this case the parameter $a$ can be varied by simple
rescaling. We investigated these solutions but, having great
difficulties in selecting the proper boundary condiditions, we abandoned
that approach.

Using the new equation of state, in (1.5), and the natural boundary
conditions (regularity at the center and fall-off at infinity) we
re-calculated the static configurations.  With $n = 3$ (inside) and $n =
6$ (outside), solutions were found for values of the parameter
$a$ ranging from $10^{-6}$  to 1/5.765 and no solutions were found for
larger values of $a$.   A relation
between radius and mass emerges in all cases considered, including:

(1) Polytropic index  $1\leq n\leq 3$ (near the center) and 6 (at
large  distances).

(2) Index $n = 3$  and 15.

\no All these static solutions appear to be stable to radial
perturbations.
It may be objected that the equation of state used here is somewhat
special, not
to say {\it ad hoc}. We are nevertheless justified in concluding that
instabilities of polytropes found previously are characteristic of
a restricted class of boundary conditions; they are not
generic.

An important consequence of the fact that the dynamics is formulated as
an action principle is the existence of a conserved current. With
the usual boundary conditions at the center, and natural boundary
conditions at infinity, we find that the asymptotic mass is a constant of
the motion.

Section 4.   An application to the sun predicts the central density and
pressure, close to the values obtained within the traditional approach.
In applications to white dwarfs the constant $a$, a free parameter in other
cases, is known. In the case of complete degeneracy the polytropic index
is equal to 3, and in this case a unique mass is predicted, very close
to the mass of the sun. This limit, here derived from a theory in which
all the stars are stable, is  close to the limiting value obtained by
Chandrasekhar from stability considerations [C1].

The instabilities of white dwarfs discovered by Chandrasekhar are
difficult to  interpret,
as witness the reservations expressed by Eddington. The first question
that
comes to mind is the future development of a star that starts from
static but unstable initial conditions. This could not be  answered
within  the
context in which the instabilities appeared, because that context was
not a
mathematically well defined  model. Instead, Chandrasekhar's results
have been
taken to mean that there are limitations to the range of physical
parameters
(mass and size) that are possible, given the assumed thermodynamic
properties of the star. With this conclusion our results are in perfect
agreement:  There is a largest mass, beyond which
the problem is not that the static solutions are unstable, but that they
do  not exist.

The maximal mass of a neutron star can be obtained once the critical
density is known. Using commonly accepted values we again recover the
traditional limit. We explore some variations of the representation used
for the piecewise almost constant function $n[\rho]$ and
discover an unexpected density profile.

\ve

  \no  {\steptwo 2. Matter model and equations of motion }

We add the matter model  action
$$
A_{matter} = \int d^4x\sqrt{-g}\Big({\rho\over 2}( g^{\mu\nu} \psi_{,\mu}
\psi_{,\nu} - c^2) - V[\rho]\Big) =:  \int d^4x\sqrt{-g}~{\cal
L}~
$$
to the Einstein-Hilbert action and restrict the metric to the spherically
symmetric form
    $$
(ds)^2 = {\e}^\nu(dt)^2 -  \e^\lambda(dr)^2  - r^2d\Omega^2,
$$

Einstein's equations then reduce to
$$\eqalign{&
G_t^t =  -e^{-\lambda}\Big({-\lambda'\over r}+ {1\over r^2}\Big) +
{1\over r^2} =  8\pi G \Big(\e^{-\nu}\rho\dot\psi^2 -{\cal L}\Big),
   \cr  & {G_r}^r =  -{\rm e}^{-\lambda }\Big({\nu'\over r} +
{1\over r^2}\Big)+ {1\over r^2} =  8\pi
G\Big(-e^{-\lambda}\rho(\psi')^2 - {\cal
L}\Big),\cr &G_t^r = \e^{-\lambda} {\dot \lambda\over r} =  - 8\pi
G\e^{-\lambda}\rho
\psi'\dot\psi,
\cr}\eqno(2.1-3)
$$
and are supplemented by the wave equations, from variation of the fields
$\rho$ and $\psi$,
$$\eqalign{&
{1\over 2}\Big(\e^{-\nu}\dot\psi^2-\e^{-\lambda}(\psi')^2-1\Big) =
{d V\over d\rho} , \cr
\p_t(&\e^{(-\nu+\lambda)/2}r^2\rho\dot \psi
)-(\e^{(\nu-\lambda)/2}r^2\rho\psi')'= 0.\cr}\eqno(2.4-5)
$$
The Lagrangian density ${\cal L}$, evaluated on-shell, is interpreted as
the pressure, subsequently denoted $p$. This is not only because of its
appearance in the expression for the energy-momentum tensor - in Eq.s
(2.1-3) - but also because of the fact that ${\cal L} = \rho(dV/d\rho) -
V$, which is a familiar expression for pressure in terms of internal
energy; see [FW], page 304. (Extending this to an
off-shell identification of the pressure with the Lagrangian
density would be a mistake.)

The function $\lambda$ is often replaced by the function $M$ defined
by
$$
M: = {r\over 2}(1-\e^{-\lambda}),~~ \e^{-\lambda} = 1 - {2M\over
r};\eqno(2.6)
$$
   then Eq.s (2.1-2) can be written as follows,
$$\eqalign{
M' &=  4\pi G\, r^2(\e^{-\nu}\rho\dot \psi^2-{\cal L}),\cr
r\e^{-\lambda}\nu' &= 1-\e^{-\lambda} + 8\pi G\, r^2(\e^{-\lambda}\rho
\psi'^2 + {\cal L}).
\cr}\eqno(2.7)
$$
The two equations can be combined to yield
$$
   (\nu + \lambda)'  = 8\pi G\, r\,\e^\lambda\rho\,(\e^{-\nu} \dot\psi^2
+
\e^{-\lambda}\psi'^2).\eqno(2.8)
$$
The differences between this theory and the usual phenomenological one
are mainly as follows:
\b

\no$\bullet$  Eq.(2.4) looks unfamiliar, but taking the derivative with
respect to $r$ one recovers the usual force equation with only minor
changes. In the static case, when $\dot \psi = 1, \psi' = 0$ and
$\nu,\lambda$ are time independent, this becomes the hydrostatic
condition,
$$
p'/\rho = -{1\over 2} \nu'\e^{-\nu}
$$
  and this has exactly the same form
in both theories. But Eq.(2.4) is  stronger than its derivative; it
furnishes an additional constraint on the boundary, and it is this
equation that provides a relation between the radius and the mass. In the
traditional approach the boundary of a polytrope is often chosen to be at
the point where the pressure becomes  zero,
that  always exists if the polytropic index $n =
(\gamma-1)^{-1}$ is less than 5.   There is no effective or
meaningful matching of the field $\nu$ to an external Schwarzschild
metric, a fact that, in our opinion, makes the whole proceeding
unsatisfactory. In our model this matching is  expressed
by
$$
\nu(R) + \lambda(R) = 0,~~  1-e^{\nu(R)}   ={2mG\over R}.\eqno(2.9)
$$
The first equation determines $R$ and the second gives the value of the
mass.

\b
\no$\bullet$  Another significant difference is that
the old approach  has no intrinsic conserved current (a conserved
baryonic current is often introduced by hand [C2]), while the new theory
  does, namely
$$
J^\mu =   \sqrt{-g}\,\rho\, g^{\mu\nu}\partial_\nu\psi.\eqno(2.10)
$$
The existence of this conserved quantity   does not by
itself assure us that the mass is a constant of the motion. However,
with
the new equation of state introduced later, and the natural boundary
conditions that it entails, it does indeed turn out that the mass is
conserved. (Section 3.)
\b

\no$\bullet$ In the static case the pressure of the model corresponds
exactly to the pressure as defined by Tolman's formula, while Tolman's
density is replaced by $ \rho + (\rho V)'$.
    We do not try to guess
the precise physical interpretation of $\rho$ and $p$ but try instead
to obtain results in terms of quantities that we feel sure
    are physical, such as the gravitational mass, uniquely
defined by the asymptotic gravitational potential.

In the static case, when $\dot\psi = 1$ and $\psi' = 0$, we find a
curious, special solution, with $f$ constant, namely
$$
\e^{-\nu} = 1 + 2a\gamma f,~~ \e^{-\lambda} = 1 + 8\pi G p r^2.\eqno(2.11)
$$
As a global solution it is of no interest, but we shall find
a solution for which $f$ remains nearly constant over a finite
interval.

In the paper [F2] we examined  static solutions
   with boundary conditions determined  by matching the metric of
the interior polytrope  to an exterior Schwarzshild metric. As explained
above, the radius and the mass were determined by the two
conditions in Eq.(2.9). Solutions were given  for
$n = 1,2,3,4, 6,10,...
$~, but only the case $n = 3$ will be invoked here;
   Table  1 reproduces the data for this case.

  \ve

\ce{\bf Table 1. Mass/radius relation for $n$ = 3 and exterior
Schwartzschild.}
\def\s{\scriptstyle}{\stephalf

\settabs \+ ~     & 1~~~~~  ~   ~~~~& 20~~~~~ ~ ~~~& 100~~~~~~~~&
1000~~~~~~& 10000~~~~~~&~~~~~~~~~  &
100000~~&~~~~~~~~~~~~&~~~~~~\~~~~~~~&~~~~~~~~~~~~~&~~~~~~~~~~~~&~~~~~~~~~~~&~~~~~~~~~\cr

\+&$-\s\nu(0)$&.0001&.001&.005&.02&.1&.3&.4&.45&.475& .49&.5\cr

\+ &$\s R$&18750&1877&374& 92.4& 17.4&5.03&3.59&3.12&2.90&2.84&
2.77\cr

\+ &$\s  $&60 000&6000&1200& 300&57&18.6&14.4&13.1&12.7&12.5&
12.3\cr

   \+&$-\s\nu(R)$&.0000542&.000542&.00270&.01073&.0516&.1388&.1721&.1863
&.2026&.1955&.1927\cr

\+&$\s
2mG$&1.017&1.015&1.001&.991&.895&.680&.618&.582&.567&.555&.547\cr

\+&$\s
2mG/R $ &.000054&.00054&.0027&.011&.051&.135&.172&.187&.196&.195&.197\cr

\+& $-\s \nu(0)$&.6&.7 & .8&.85&.9 & .95&1&1.2&1.3&1.5&2&2.065 \cr

\+&$\s R$&
2.30&2.03&1.96&2.022&2.17&2.43&2.91&7.86&8.99&7.64&4.82&4.66
\cr

\+&$\s  $&
11.4&11.5&136&14.4&16.8&20.5&26&48&49&47&29&28.5
\cr

   \+&
-$\s\nu(R)$&.217&.232&.2012& .1875&.1688
&.1458&.120&.0524&.05142&.0650&.1000&.1021 \cr

\+&$\s 2mG$&.489&.441&.396&.379&.365&.354
&.350&.404&.450&.497&.482&476
\cr

\+&$\s
2mG/R$&.213&.217&.202&.197&.168&.146&.120&.051&.050&.065&.100&.102\cr}
 \vskip.5cm 

There appears to be an upper limit to the mass of about 1.02, and a
lower
limit on the radius of about 1.96, in units where $c = G = a = 1$.
The number in the third row is the position of the first zero of the
pressure. Since the scale is not fixed the most significant data are
the range of the dimensionless ratio $2mG/R$ and the correlation of this
number with  $\nu(0)$. This is shown in Fig.1, lower curve.
\b
\ce{\bf Oscillations around the static solutions}

In this section we use units such that $ 8\pi G
=1$ and fix the scale by setting $a = 1$. We linearize the equations at
the
static solution. The equation for
$G_r^r$ gives
$\delta p$,
$$
   r \e^\lambda \delta p =    \delta \nu'
-( \nu'   + {1\over r })\delta\lambda ,\eqno(2.11)
$$
and so does ``Newton's equation", Eq.(2.4)
$$
(\e^{-\nu/2} ) \delta(\e^{-\nu/2}\dot\psi ) =  \delta{dV\over
d\rho} = { 1\over \rho}\delta p,
$$
or
$$
\delta p = \e^{-\nu }\rho  ( \delta\dot\psi-  { \delta\nu/2}
) .\eqno(2.12)
$$
    Eliminating
$\delta p$ from these two gives
$$
   r\rho\e^{\lambda-\nu }  ( \delta\dot\psi-  { \delta\nu/2}
) =\delta \nu' - ( \nu'  + {1\over r })\delta\lambda
.\eqno(2.13)
$$
Equation (2.8) for $G_t^t-G_r^r$ gives $\delta\rho$,
$$
   r\e^{  \lambda-\nu} \Big(\delta\rho + \rho  \delta\lambda
+ 2\rho(\delta\dot\psi-\delta\nu/2)\Big) = \delta \nu'
    + \delta\lambda',
$$
or 
$$
   r \delta  \rho  =  -2 r\rho
   ( \delta\dot\psi-   { \delta\nu/2} ) -
r\rho \delta\lambda
    +\e^{\nu -\lambda} ( \delta \nu'
    + \delta\lambda'  ) \eqno(2.14)
$$

\ve

\parindent=1pc

   \vskip1.1in
\def\picture
#1 by #2 (#3){
   \vbox to #2{
     \hrule width #1 height 0pt depth
0pt
     \vfill
     \special{picture #3} 
interface
     }
   }
\def\scaledpicture #1 by #2 (#3 scaled #4){{

\dimen0=#1 \dimen1=#2
   \divide\dimen0 by 1000 \multiply\dimen0 by
#4
   \divide\dimen1 by 1000 \multiply\dimen1 by #4
   \picture \dimen0
by \dimen1 (#3 scaled #4)}
   }

\parindent=1pc

\vskip-3cm
\epsfxsize.7\hsize
\centerline{\epsfbox{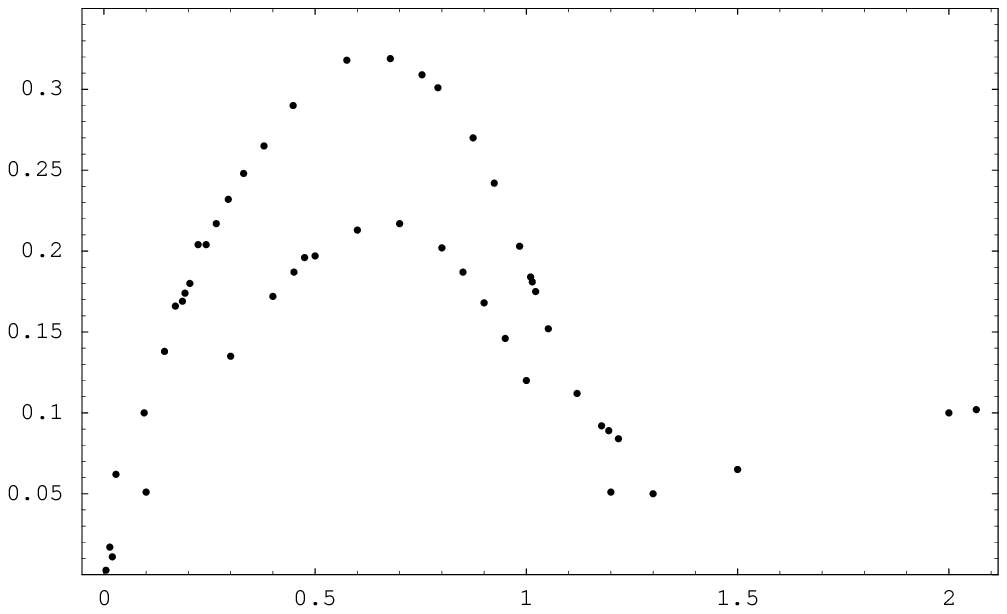}}
\vskip-1cm
\vskip1cm
\no{\it 
Fig.1.  The lower curve shows the essential information from Table 1. The abscissa is $\nu(0)$
and the ordinate is the dimensionless ratio $2mG/R$, where $R$ is the  boundary defined by Eq.(2.9).
The upper curve shows the same information from Table 2, the ordinate is here $2mG/R_{cr}$.} 
\b
Using (2.13) we eliminate the time derivative and obtain a
constraint,
   $$
   r\delta  \rho  =
   \e^{\nu -\lambda}\Big(\delta\lambda'- \delta \nu'
      +  ( \nu'-\lambda'   + {2\over
r })\delta\lambda\Big)\eqno(2.15)
$$

\b

The
original set of equations is equivalent to (2.12) and (2.15) that
determine
$\delta p$ and
$\delta
\rho$, and the two equations
$$
   r\rho\e^{\lambda-\nu } ( \delta\dot\psi- { \delta\nu/2} )
=\delta \nu' - ( \nu'  + {1\over r })\delta\lambda
   \eqno(2.16)
$$
and
$$
\delta \dot\lambda =  r\rho \psi' \eqno(2.17)
$$
for the functions $\lambda$ anf $\psi$.
The only difficulty is that there is no determination of $\delta\nu$;
to fix it we have to use the equation of state; the choice may
affect stability.

\b\ce{\bf Equation of state }

We express $V$ and $\rho$ parametrically, in terms of the Emden
function $f$:
$$
V = af\rho,~~\rho = f^n,~~ p = \rho{dV\over d\rho} V,
$$
with
$$
n = n_1 + {n_2-n_1\over 1 +  f^K}.
$$
This function takes the constant value $n_1$ (in most of our
calculations $n_1 = 3$) in the inner regions where $f > 1$, and the
constant value $n_2$ (the values 6 and 15 were explored) in the
atmosphere. The critical region is thus at the place where
$f$ takes the value unity; $f$ and $\rho$ are entered as multiples
of their critical values. This choice, with $a$ constant, gives the
best approximation to an isentropic equation of state, as we shall see.

In regions where
$n$ is constant we have
$p = (a/n)f^{n+1}$. In the boundary region  there will of course be
deviations from this equation of state. We calculate
$$
  {dV\over d\rho} =  af{1+n-N\ln f\over n -N\ln  f},~~ N =
{(n_1-n)(n-n_2)\over n_1-n_2}K.
$$
Because of the high value of $K$ that was used ($K$ = 50), the function
$N$ vanishes except in a narrow interval where $n_1<n<n_2$. Of course,
because
$K$ is large, $N$ is not small in this region. However, for the same
reason, the  function $f$ varies little from its critical value in this
interval. Therefore, taking this critical value to be unity, making $f
\approx 1 $ in the interval, is highly beneficial. The two
  functions
$
\rho{d \over d\rho}V, ~~\gamma V.
$
are nearly indistinguishable over the entire interval $0<r<\infty$.
  The deviation is small, in a
small region, and it will be ignored in  the calculations. (Most
results show a remarkable insensitivity to details of the equation of
state near the critical point.) Thus we set $p = a\rho^\gamma$ and
$$
  \rho\delta p  = \gamma p \delta \rho. \eqno(2.18)
$$
   For a stationary solution, when $\delta\dot\lambda
= s\delta\lambda$ for some number $s$ that we hope will have to be pure
imaginary, from (2.11),(2.15) and (2.18),
$$
\delta \nu'= r \e^\lambda \delta p
+( \nu'   + {1\over r })\delta\lambda ,
$$
   $$
   r\rho \psi' = s\,\delta \lambda ,
$$
$$
\delta\lambda' =rK\e^\lambda    \delta p +
    (\lambda'-{1\over r})\delta\lambda,~~ K = 1 +
\e^{-\nu}\rho/\gamma p.
$$

Elimination of $\delta p$ leads to
$$
\delta \nu'= r \e^{\lambda -\nu}\rho( s\delta\psi - \delta\nu/2)
+( \nu'   + {1\over r })\delta\lambda ,
$$
   $$
   r\rho \psi' = s\,\delta \lambda ,
$$
$$
\delta\lambda' =sr\rho K\e^{\lambda-\nu} \delta\psi - (r\rho K/2)
   \e^{\lambda-\nu} \delta\nu  + (\lambda'-{1\over r})\delta\lambda.
$$
 
\no or
   $$
(d/dr)\pmatrix{\delta\nu\cr \delta\psi\cr \delta\lambda\cr}
=  \pmatrix{-(r \rho/2)\e^{\lambda-\nu}& s(r \rho )\e^{\lambda-\nu}&
   \nu' + {1\over r} \cr 0&0&(s/r\rho) \cr
- (r\rho K/2)\e^{\lambda-\nu} &s(r\rho) K\e^{\lambda-\nu}  &
\lambda'-{1\over r}\cr}\pmatrix{\delta\nu\cr \delta\psi\cr
\delta\lambda\cr}.\eqno(2.19)
$$
  From this it is easy to see that the integration from $r = 0$ can
proceed, we
start at $r = 10^{-10}$ with $\delta\psi = \delta\lambda = 0$ and
$\delta\nu\neq 0$.  This will make $\delta\nu'$ and $\delta\lambda'$ of
order $r$, $\delta\lambda$ of order $r^2$ and $\delta\nu -
\delta\nu(0)$
of order $r^2$.

The solutions include a simple gauge transformation; when it is ignored
the
system can be reduced to a single, second order differential equation
for the
function
$L = r\e^{-\lambda}\delta\lambda$:
$$
\ddot L ={r\rho\over 2}(\nu' + {1\over r})L +
r^2\rho\,\e^{-(\nu+3\lambda)/2}\big({\e^{(\lambda+3\nu)/2}L'\over
r^2\rho K
   }\big)'.\eqno(2.20)
$$
This equation shows that the stationary solutions have frequencies
determined by a self adjoint Sturm-Liouville operator, with a domain 
of functions $L$ that satisfy a condition that fixes the value
of
$L'/L$ at the boundary and such that
$L/r^3$ is regular at the origin.

\b

\ce{\bf Boundary conditions, difficulties}
Round 1. Here we report the result of studying the stability of the static
solutions in the case that the star has a fixed polytropic index in the
interior and the metric is matched to an exterior Schwarzschild metric
at the value of $r$ determined by Eq.s (2.9). (These are  the solutions
listed in Table 1.) We adopt Eddington's boundary conditions at the center
[E], so that the solutions of the  static equations of motion for fixed
values of
$n$ are indexed by  the value $\nu(0)$ of the function $\nu$ at the
center. Matching of the solution to the exterior metric -
Eq.s (2.9) - determines both the radius and the gravitational mass. The
result is a relation between mass and radius, for each value of $n$,
reproduced for the case $n = 3$ in Table 1 and in Fig.1, lower curve.

The stability of a star   is to be determined by
solving
  equations (2.19) or (2.20) with some boundary conditions. It is
difficult, however, to understand what boundary conditions are
appropriate.

To deal with  Eq.(2.20) from the point of view of Sturm-Liouville
theory, one looks for solutions with harmonic time dependence,
$L(x,t) =
\e^{i\omega t}L(x,0)$. One identifies the range of the parameter
$\omega$ with the spectrum of an operator in a
Hilbert space constructed from a space of functions of $r$. An
acceptable
set of boundary conditions must make this operator self adjoint.

It has always been assumed that the origin is a regular point.
It is difficult to find a real justification for this, since the
center of the star is a region about which one has very
   little
information. Nevertheless,  we follow this precedent since it
helps to give a precise mathematical sense to our model. Thus
$\nu$ has a well defined value at   $r = 0$,   $\lambda$ is of the
order of $r^2$ and the function $L$ is of order $r^3$. In this case the
possible additional boundary conditions amount to fixing  the value
of $L'(r)/L(r)$ at the surface. If we fix the boundary and suppose that $L
= 0$ there, then we find a discrete set of oscillating solutions and, in
  some cases, decaying solutions, in agreement with the findings of
Chandrasekhar [C1]. But the radius of the star is determined by the first of
the conditions (2.9), and that means that the surface of the star is
pulsating, as we have verified numerically. This seems not at all
unnatural, the difficulty is that the asymptotic mass, as determined by
the matching of the metrics, is also pulsating. We are very skeptical of
these results. In the traditional treatment mass has no
relation to a conserved quantity, and an uncertain relation to the
asymptotic mass; this may be the reason why the problem has  not been
addressed in connection with the work of Chandrasekhar.

Round 2. Fixing the mass would seem to be the more
reasonable boundary condition, since it is determined asymptotically in
a region where the density is zero. It does not seem possible that
oscillations of a finite star propagate to infinity through an infinite
region of empty space. To understand this better we should remember
that the external Schwarzschild metric, according to our interpretation
[F2], is not the  metric
of empty space, but a singular limit of a family of metrics of spaces
with non vanishing density. Consequently, in our next attempt we
replaced the empty Schwarzschild exterior by a crude approximation for
the atmosphere, another polytrope. The mass distribution now extends to
infinity, though the density falls off extremely rapidly and the metric
soon becomes indistinguishable from that of Schwarzschild. The asymptotic
mass is  a property of the exterior polytrope, but we were unable to
match the two polytropes at the boundary in such a way that this mass
would remain constant. Still we cannot claim to have excluded this
possiblity completely.

It has been traditional since the beginning, to admit a discontinuous
behaviour of density and pressure at the surface of a star. This is
reasonable if the star is cold, but perhaps less likely to be
typical of a polytrope.  Facing doubts of this kind, and the
difficulties discussed in the preceding paragraph, we came to the
realization that it may be better to give up the idea of a fixed
boundary  and introduce the  equation of state described in the
introduction, thus allowing all the fields to vary continuously
throughout.

  \b

\bb
  \no{\steptwo 3.  Improved boundary conditions}

Round 3. We suppose that there is a region of critical density $\rho_{cr}$,
where the polytropic index changes more or less abruptly from a value
$n_1<5$ (in our calculations  $n_1\leq 3$) that is appropriate for the
bulk of  the
star, to a value $n_2>5$   (actually 6 or 15) that we hope may be
appropriate for the atmosphere. Precisely,
$$
V[\rho] = a\rho_{cr}  \tilde  \rho ^\gamma,~~ \gamma = 1 + {1\over
n},~~ n = {n_1\tilde f^K + n_2  \over 1 +\tilde f^K}~,\eqno(3.1 )
$$
where  $K$ is a suitably large number (actually 50) and $\tilde f =
f/f_{cr}, \tilde\rho = \rho/\rho_{cr}$ are all
dimensionless. The critical density now appears as a common factor of the
energy momentum tensor, and on the right hand sides of Eq.(2.1-3),
together with
$G$.   We shall drop the tildes on $\tilde f$ and $\tilde\rho$,
   so that $f$ and $\rho$  are henceforth  given as multiples of
their critical values. Then Eq.s (2.1-5) remain valid if the factor $G$ in
(2.1-3) is replaced by $G\rho_{cr}$. Finally we choose our unit of length
such that this factor is equal to unity,
$$
G\rho_{cr} = 1.
$$
The pressure in these units is $  (a/n) \tilde \rho^\gamma$ and the
critical pressure is
$(a/  \bar n)\rho_{cr},~
\overline n  = (n_1 + n_2)/2$.

   There  is no longer any question of
matching to an exterior Schwarzschild metric,  instead we require that
the metric
   approach the Schwarzschild
form at
large distances, to order $1/r$. The mass is determined by this asymptotic
metric,
$$
2mG := \lim r \lambda(r)  = -\lim r \nu(r).
$$
In the traditional approach the second condition is not effective,
since only the derivative
$\nu'$ of the function $\nu$  appears
in Einstein's equation.
The extra condition that comes from the action
principle guarantees that the metric is asymptotically Schwarzschild (as
  $r^3\rho(r) \rightarrow 0$) so that the two limits always coincide.
The indices $n_1,n_2$ were given the values 3,6 in our initial
calculations, Table 2. A larger value of $n_2$ makes the metric approach
more quickly to the
Schwarzschild form, Table 3. The exponent $K$ determines the abruptness of
the  change of
the index from 3 to 6, our calculations were done with $K = 50$.

Since the index is not constant,  the
value of
$a$ can no longer be reduced to unity by a change of scale.

As before, we assume that all the fields are regular
at the origin.  A star is characterized by the parameters $f_{cr}$ and
$a$.   We choose a value of $a$ and determine allowed
value(s) of

\no$\nu(0)$ by demanding that $-r\nu(r)$ tend to a finite limit (twice the
mass $m$) at infinity, and establish in this way a correlation between
$R$ and $m$.  Results are given in Tables 2-7. The upper curve in Fig.1
shows the dimensionless number
$2mG/R_{cr}$ versus $\nu(0)$, for the case $n = 3,6$ (3 inside and 6 outside).  The
lower curve was obtained by matching the interior solution ($n$ = 3) to  an
exterior Schwarzschild metric. The difference between the two curves is easily
explained since the upper curve refers to the critical radius $R_{cr}$
while the lower curve is $2mG/R$, where $R$ is the boundary.
\b
\ce{\bf All static solutions are stable!}

When there
is no fixed boundary, and the star extends to infinity, the asymptotic
behaviour becomes important; for the function $L$ that embodies the
oscillations around a static solution we find
$$
L \propto  \sin(r^{3/2}b)/r^k
$$
with $k$ and $b$ constant. Values of the exponent $k$ determined by
numerical calculations for $n = 6$ ($ k = 5/2$) and $n = 10$ ($k =
9/2$)  at
large distances  are such that the metric fluctuations $\delta\nu$ and
$\delta\lambda$ fall of faster than $1/r$. It follows that the mass
(defined
by the asymptotic metric) is unaffected by the oscillations.

With the improved boundary conditions the real spectrum of frequencies
is
continuous and apparently the entire real line. No unstable solutions
were
found. This could have been anticipated by inspection of Eq.(2.20). The
factor
$\rho$ in the first term on the right hand side makes this term fall
off very
fast at infinity, which suggests that this term cannot affect the
  spectrum of $\omega^2$. The conclusion must be that the instabilities
first discovered by Chandrasekhar [C1] are imposed on the theory by the
choice of boundary conditions.
\b
 
\ce{\bf Constants of the motion!}

The  conservation law (2.5) can be integrated to yield
$$
{d\over dt}\int_0^\infty\sqrt{\e^{(-\nu + \lambda)/2}} r^2 \rho\dot\psi
dr =
   \Big[\sqrt{\e^{(\nu-\lambda)/2}} r^2\rho\psi'\Big]_0^\infty.
$$
   In view of the boundary conditions at the origin,
$$
{d\over dt}\int_0^\infty\sqrt{\e^{(-\nu + \lambda)/2}}   \rho\dot\psi
r^2dr  =
    \lim_{r\rightarrow \infty}\Big[\sqrt{\e^{(\nu-\lambda)/2}}
r^2\rho\psi'\Big].
$$
The factor $\rho$ on the right hand side suggests that there is no flux
at
infinity, but in fact the flux $ r\rho\psi'$ is   equal to
$-\delta\dot \lambda/8\pi $ by Eq.(2.3). For a static configuration both
sides  of this
equation are zero; for a first order deviation from a static
configuration we have
$$
{d\over dt}\int \sqrt{\e^{(-\nu + \lambda)/2}} r^2
\rho\delta\dot\psi
drd\Omega = {1\over 2}
   \lim_{r\rightarrow \infty}\Big[\sqrt{\e^{(\nu-\lambda)/2}}
r\delta\dot\lambda\Big].
$$
If the perturbed and unperturbed metrics both tend  to Schwarzschild at
infinity, then
$r \delta\dot\lambda \rightarrow 2 \dot  m$ so that, finally,
$$
{d\over dt}\int \sqrt{\e^{(-\nu + \lambda)/2}} r^2
\rho\,\delta\dot\psi
drd\Omega =  dm/dt .\eqno(3.2)
$$
It is not {\it a priori} obvious that the asymptotic mass  is  a
constant
of the motion, but a result of our calculations  is that
$r\delta\lambda$
tends to zero at infinity so that in fact $\dot m = 0$. The
asymptotic mass is a constant of the motion and so is the quantity
$$
  \int d^4x \sqrt{-g}\,g^{tt}\rho
$$
In our numerical study this number turns out to be bounded upwards by the
asymptotic mass, the difference being greater in the case of strong
gravitational fields, which suggests an interpretation of the difference
as a binding energy. This is the interpretation suggested in [KW], in the
context of the traditional treatment, with the difference that they
replace $\rho$ by $\hat \rho$ (Tolman's density). But the
integral of $\hat \rho$ is not a constant of the motion!
\bb

\ve
\no{\steptwo 4. Applications }

\ce{\bf Modelling the sun}

The results for $n_1 = 3, n_2 = 6$ (Table 2) and for $n_1 = 3/2, n_2 = 6$
(Table 6) are shown in in Fig.2. The graph of allowed values
of
$(R_{cr},2mG)$ has a lower branch with small internal pressure, a maximal
value of
$R_{cr}$, and an upper branch with increasing central pressure and a
maximum value of the mass.

 \b
\epsfxsize1.0\hsize
  \centerline{\epsfbox{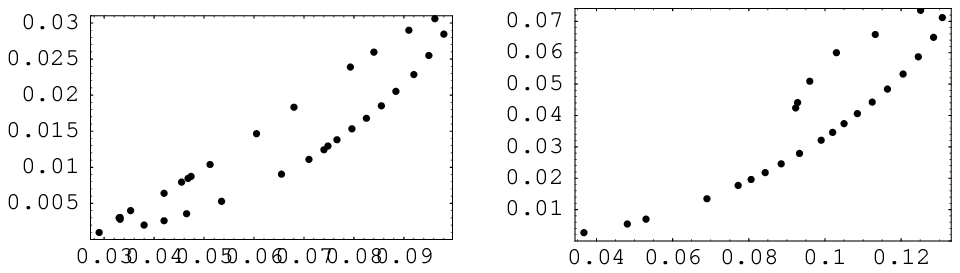}}
\
\topskip-1in
\vskip-.5cm

\no{\it 
Fig.2.   The mass-radius relation obtained with different boundary equations of state.
On the left $n = 3,6$, on the right $n = 3/2, 6$. }
\b
The bulk of the sun is often modelled with a polytrope with index 3. The
parameter values are
$$
(2MG)_{sun} = 2.95\times 10^5 cm,~~ R_{sun} = 6.96 \times10^{10}cm,
$$
 and the ratio is $.424 \times 10^{-5}$.  We must not equate $R_{sun}$
with the value $R_{cr}$ of the radius at which the density takes the
critical value; this may occur deep in the interior.

If the critical density is $\rho_{cr} =  k^2 g/cm^3$, then
$$
\rho_{cr}G = .7414\times 10^{-28} k^2 /cm^2
$$
and the unit of length used
in our calculations is therefore
$$
\ell = {1.16\over k}  \times 10^{14} cm.
$$
Here are 2 examples from Table 2.
\b
Example 1. Take $1/a = 10^{5}$, the Table gives
$
2mG = .32\times 10^{-7} ~\ell =   37 \,k^{-1}\times 10^5 $.
To agree with the sun value we need $k = 37/2.95 = 12.54$. Thus $\rho_{cr}
= 158~ g/cm^3$, which may be a little high. The  critical radius is
$9.23\times 10^{-4} \ell =  .854\times 10^{10}$ and the critical pressure
is $p_{cr} = (a/3)\rho_{cr} = .527\times 10^{-3}$ or $4.74\times 10^{17}
dyn/cm^2$.
\b

  Example 2.  Take $1/a = 2\times 10^{5}$, the Table gives
$
2mG = .113 \times 10^{-7} ~\ell =  .131 \,k^{-1}\times 10^7 $.
Here we need $k = 13.1/2.95 = 4.44$. Thus
$
\rho(0) = 1.257\times \rho_{cr} = 24.8 ~g/cm^3
$, which may be too low.
  The critical radius is
$$
R_{cr} = 6.52\times 10^{-4} \ell =
1.7\times 10^{10}.
$$
This is about $R_{sun}/4$. The density profile, shown in Fig. 3, shows
that this is quite reasonable. The pressure is
$$
p(0) = {a\over 3} \rho_{cr} f(0)^4 = .446  \times 10^{-4},
$$
or $.401 \times 10^{17} dyn/cm^2$. 
\vskip.5in
\epsfxsize.78\hsize
  \centerline{\epsfbox{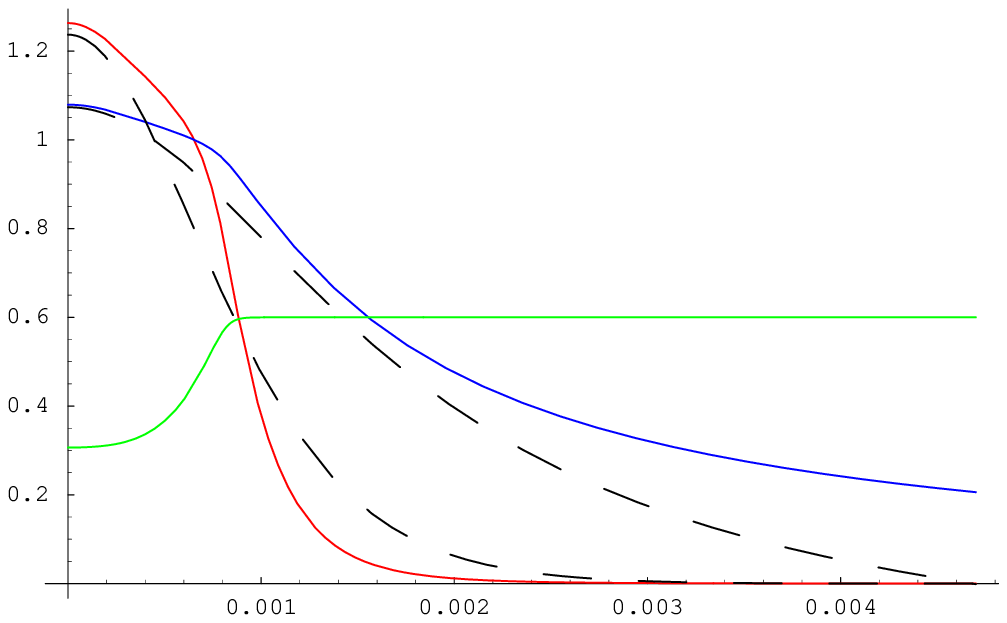}}
\
\topskip-1in
\vskip0in 
\no{\it 
Fig.3.  Various functions plotted against the distance from the center. The function with the step is 
n(r)/10. The other solid curves show $f(r) $ and $\rho(r)$. The dashed curves show analogous results with a polytropic index fixed at the value 3. What cannot be seen is that the latter pass to negative values at $r = .0047$, while the solid curves never do.} 
\b

Given the known values of the mass
and the radius of the sun, the traditional polytrope model predicts the
values $\rho(0) = 76.39 g/cm^3$ and $p(0) = 1.24 \times 10^{17} dyn/cm^3$
for the
   density and pressure at the center.     Our dynamical
model gives slightly different values
for the central density and pressure. Various refinements, such as
larger values of $K$ and/or $n_2$, may affect these predictions to a
limited extent.

Let us compare the effect of different boundary conditions for fixed
values of $\nu(0)$ and $a$. In the case of Example 2, in units of
$10^{10} cm$:

$\bullet$ The Emden function vanishes at 12.3. This would be the
prediction for the radius of the standard approach with this choice of
parameters (not the best choice). The density predicted by the model at
this point is about 1/5000 of the critical value.

$\bullet$ The actual value of the visual radius is 6.96.

$\bullet$ Matching to an external Schwarzschild metric would fix  the
boundary at 3.9. The density at this point is about 1/8 of the critical
value, dropping  abruptly to zero.

$\bullet$ The critical radius is at 1.7.

$\bullet$ The density profile is shown in Fig. 3, and the density
predicted by the standard theory is shown for comparison.

All three models can be made to give the correct radius and mass of the
sun. They differ, but not very greatly, in the predictions for central
density and pressure, and they differ considerably in the degree to which
they seem to be physically reasonable.

Fig.4 shows profiles of the metric functions.

\vskip.5in
\epsfxsize.78\hsize
  \centerline{\epsfbox{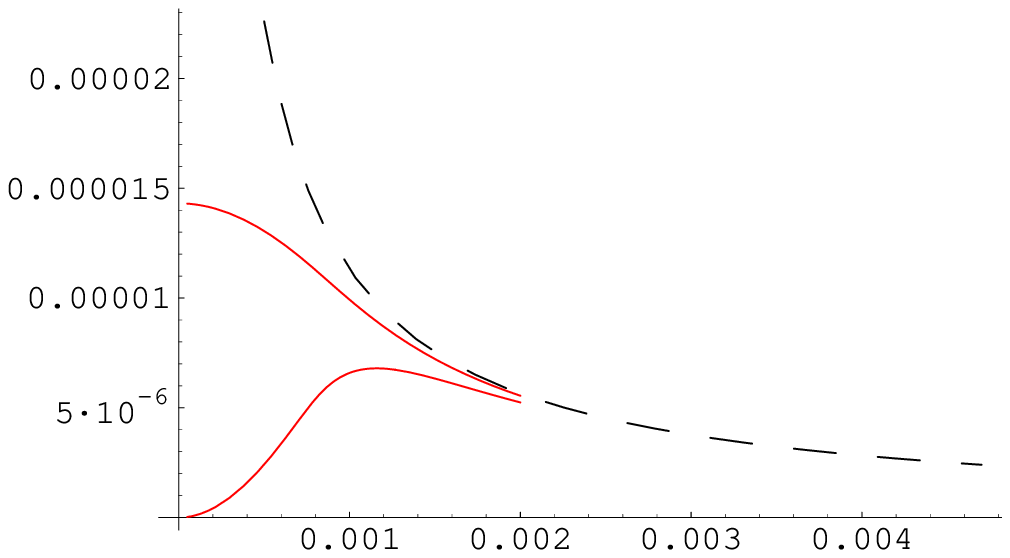}}
\
\topskip-1in
\vskip0in
\no{\it 
Fig.4.    The metric functions $\lambda$ (lower curve)  and $-\nu$
in the case that best fits the sun.  The dashed line is the function $-\nu$ for the Schwarzschild metric with the same mass.}

\b
\ce{\bf Strong gravitational fields}

A good indicator of the strength of the gravitational field is the
dimensionless ratio $2mG/R_{cr}$. The maximum values are, for the case
that $n_2 = 6$:

$$
\hskip3mmn_1 ~~~~   =~~ ~ 3~~~~~ 5/2~~~\,~ 2~~~\, ~3/2~~\, ~~1 ~~~~1/10
$$
$$
{2mG\over R_{cr}} ~~= ~~.33 ~~~~.43~~~~.50~~~.59~~~~.67~~~~.77
$$
As the value of $n_2$ is increased the falling off of the density
outside the critical radius becomes much more rapid. Nevertheless,
there is always a region where the gravitational field is less strong
and where the density is not yet negligible. Thus it is difficult to
imagine a situation where no radiation escapes. A much simpler
assumption to explain the astrophysical ``black holes" would be for the
star, or at least the atmosphere, to be non-radiating on account of the
temperature being very low. But in the polytropic models the temperature
falls off much more slowly than the density.

\ve
\ce{\bf The Chandrasekhar limit}

Under conditions that are believed to prevail in a white dwarf the value
of the parameter $a$ can be calculated. Normally the equation of state is
expressed as $p = K\rho^\gamma$, with pressure and density given in
units of $g/cm^3$. In that case one obtains
$$
K = .548\times 10^{-6},~~ a = 3f_{cr} K.
$$
Our unit of length is
$$
\ell = {1.16\over f_{cr}^{3/2}}\times 10^{14} cm.
$$
The numbers posted for $2mG$, to be expressed in centimeters, must be
multiplied by $\ell$, including the factor $f_{cr}^{-3/2} \propto
a^{-3/2}$. To find the maximum value of the mass we have evaluated the
product of  $2mG$ in the table by this factor, with the result that it is
essentially constant at 1.01 in the upper half of the table and
eventually decreases  to about .33 near the bottom of the table. The model
is thus in agreement with the traditional treatment, in predicting a
unique mass in the case of weak gravitational fields, and a maximum value
of the mass that is reached roughly in the interval $ 0 < a < .01$.
This maximum value is
$$
2mG\,\ell = 1.16\times 1.01 \times 10^{14} (3K)^{3/2} = .84
(2mG)_{sun}.
$$
When the mass is close to this upper limit the ratio
$$
2mG/R_{cr} < .334.
$$
The maximum value is 788 times the value of $2mG/R_{sun}$, so it would
appear that a white dwarf of this mass may have a radius as little as
1/100 th that of the sun.

The model makes these predictions, basically the same as the theory of
Chandrasekhar, but without invoking instabilities. All the stars
described by the model are stable. Field profiles are shown in Fig.s 3
and 4.
\b
\ce{\bf Neutron stars}

According to Oppenheimer and Volkoff [OV], a model of a neutron star must have
a very rigid equation of state. We tried $n= 1/10$ in the bulk and $n =
6$ outside and in this way we reach the higher value of .77 for the ratio
$2mG/R_{cr}$. The model
does not furnish the scale; that is, in our case, the value of
$\rho_{cr}$. It is usual to place the central density of a neutron star
between
$10^{14}$ and $10^{15} g/cm^2$. Taking $\rho_{cr} = 10^{14}g/cm^2$ gives
us the scale $\ell = 1.16 \times 10^7$. Two sample solutions are
$$
a= 1/2, ~~2mG = 9 \times 10^5 cm,~~ R_{cr} = 17~ km,
$$
which is 3 solar masses and $2.4 \times 10^{-5}$ solar radii
and
$$
a = 4/3,~~ 2mG = 10^6,~~ R_{cr} = 13~ km.
$$
If the density is 9 times greater the numbers will be 3 times smaller.
All these numbers are consistent with current estimates.

It was found that the formula that was used for the variable
polytropic index fails, in this case of a low internal value, to
give a constant value in the interior; the value is close to $n_1 
 $ near the center, but  begins to increase at about half the
critical radius. The natural remedy would be to increase the value of the
exponent
$K$, but in that case the power of Mathematica to handle
very large numbers is overtaxed. An alternative formula, namely
$$
n[f] = {n_1 + n_2\over 2} + {n_1 - n_2\over 2}{f-1\over \sqrt{(f-1)^2 +
\epsilon}}\eqno(4.1)
$$
approaches a step function very well if $\epsilon$ is very small, and for
$3\geq n_1\geq 1$ this formula reproduces the same relation between mass
and radius as the one used to construct the tables, to a very good
accuracy. However, the density profile is affected in an interesting way
as $
\epsilon$ is decreased. In an intermediary range of the radial variable
the solution slips into the solution with constant density described in
Section 2. In this interval $\rho \approx\rho_{cr}$ and the polytropic
index is determined by the metric, more precisely by $\nu$, through
equation (2.11). This phenomenon is specific to the model, to the use of a
variational principle. Consequently, we  succeed in fixing $n=n_1$ in
the region interior to this platform only.

As shown in Fig. 5, the effect is not very important in the case $n_1 =
3, n_2 = 6$, but  for  \break smaller values of $n_1$ it is decisive. Fig. 6   shows the
density profile  for the case
$n_1 = 2,n_2 = 6$ and $\epsilon = 10^{-6}$.

Thus it seems that attempts to fix $n$ at a very low value leads
instead to a stratification of the star, with a middle region in which
the density is constant. This has an uncanny similarity with the
accepted picture of neutron stars. If the value of $n_1$ is decreased
below .3 or .2 this middle region reaches the center, and the
theoretically uncertain core shrinks to nothing. We do not have the
temerity to suggest that this picture of a neutron star corresponds to
reality, but we find it fascinating.

\bb
\ce{\steptwo 4.  Discussion of the results}

1. The success of the polytropic equation of state  in accounting for a
wide range of astronomical objects is well known and  almost
miraculous. One may feel, nevertheless, that there is some question about
the best choice of boundary conditions. There is also something a little
unsatisfactory about the definition of mass and mass density. We have
advocated the use of a variational principle, and we have found that
there is a simple and natural matter lagrangian that allows us to
reproduce all of the results of the phenomenological approach. It also
provides a conserved quantity, something that is often added as an
additional ingredient to Tolmans theory.  The new equation of state,
that interpolates between an interior polytrope and an exterior polytrope
that extends to infinity, is  justified on physical grounds; see the
Introduction. It wipes out all the uncertainties that are presented by
the more traditional approach using
   a fixed boundary [E][C1], and it allows to establish the existence of a
constant of the motion  related to the conserved
current. The idea of a double polytrope is not new, for Chandrasekhar  has
proposed a very similar equation of state [C2],[OV].    With this
equation of state all the examined double polytropes  are stable, which
is an extremely satisfactory resolution of the paradox presented by
Chandrasekhar's limit. It is worth emphasizing that the only practical
test of Chandrasekhars prediction is the observation of an upper limit
on the mass; this is a prediction of our model as well.
\ve

\vskip .5in
\epsfxsize1.0\hsize
 \centerline{\epsfbox{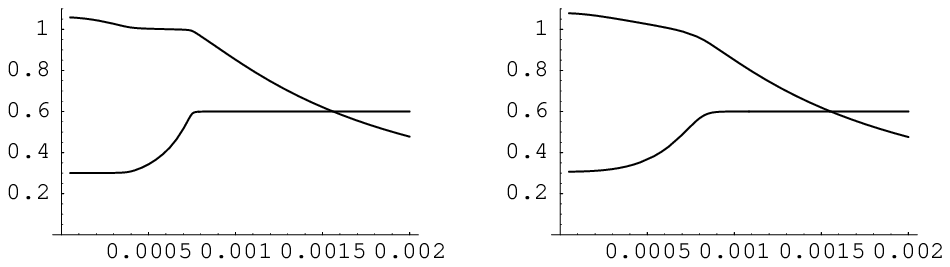}}
\
\topskip-1in
\vskip-.3cm

\no{\it 
Fig.5.   Contrast between two ways of smoothing the transition at the critical density. On the left the formula (3.1), on the right ( 4.1). In both cases, the parameters used are those that best fit the sun. } 
 
\b 
2. In this paper it is taken for granted that the metric approaches the
Schwarzschild form at infinity; in particular that the function
$r\nu(r)$
has a limit at infinity. What is called ``the mass", and denoted $m$,
is
defined by
$$
2mG = -\lim_{r\rightarrow\infty} r\nu(r).
$$
The traditional approach defines ``the mass" as the integral (in
traditional
notation)
$$
   \int \hat \rho(r) r^2 dr d\Omega;
$$
its value is the same as the asymptotic mass.
      Of all the textbooks that we
have consulted only one ([KW], pages 12-13) expresses any
discomfiture  with
this formula; they say that it is ``treacherous",  because the measure
is
wrong, given that $\rho$ is a scalar field. The  difficulty in the
traditional theory arises because it does not have a conserved current,
ultimately because it is not formulated as an action principle.
Our model is, and we have shown that,   with the  boundary conditions
that
we have adopted, there is a conserved quantity,
$$
   {1\over 2}\int_0^\infty\sqrt {-g}\,  g^{tt}
\rho \dot\psi r^2 dr d\Omega.
$$
The value is always less than $m$ and the difference can perhaps be
interpreted as a binding energy, see
[KW] page 387. To this we must say that the only meaningful concept of
energy in General Relativity is the ADM energy; therefore we make no
suggestion as to the proper name for the above integral. The
important point is that both it and the asymptotic mass are constants of
the motion. It is our opinion that this solves a  difficulty in the
applications of General Relativity to mass distributions in general and
to the structure of stars in particular.

 \vskip.5in
\epsfxsize.7 \hsize
 \centerline{\epsfbox{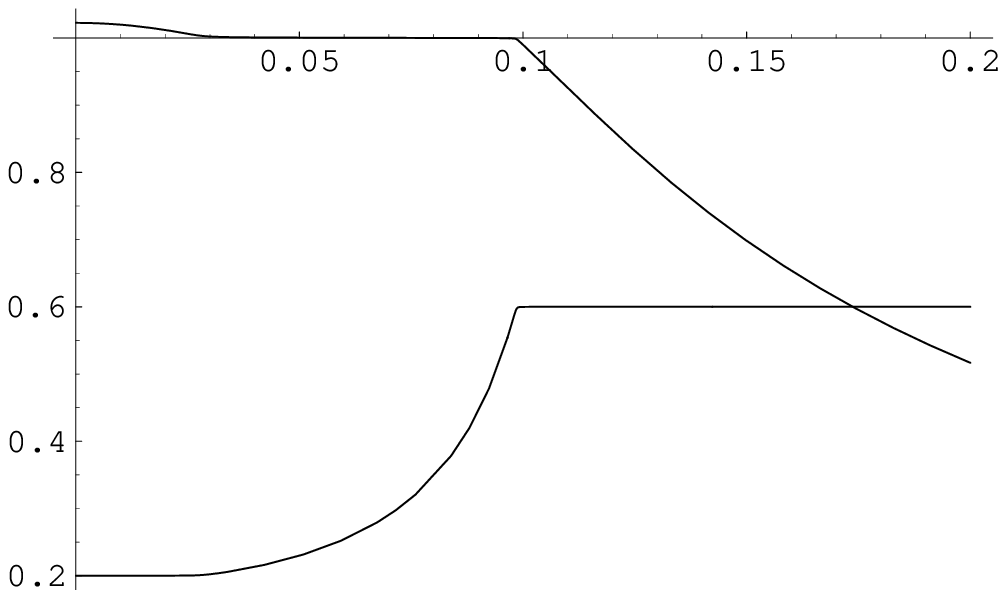}}
\
\topskip-1in
\vskip0in
\no{\it 
Fig.6.    Here the formula (4.1) was used with $n =  2,6$.  The platform of constant density reaches almost to the center. The lower curve is the function $n(r)/10$.} 

\b 

2. The paradox that is presented by the Chandrasekhar limit is that his
theory fails to explain what happens to an unstable configuration. Our
model reproduces the limits, but the unstable configurations of
Chandrasekhar's theory do  not have counterparts in the model. The
problem of explaining or predicting what happens to them simply does not
arise.

3. Observation of most real stars depends on the fact that they are
luminous.
The role of radiation in determining the equation of state has been
overlooked in this paper and this is an  obstacle to further
phenomenological
applications. The question of how radiation is to be included in the
model
is interesting. Of course, matter has to interact with radiation,
but since the interior of a star is neutral on a large scale this may
not be
the first item to take up. Radiation can be
introduced by
adding the Maxwell action to the gravitational and matter actions, but
this
is probably not the best way. Here we suggest, but offer no
justification for
it, to include  a contribution to the action of the form
$$
A_{\rm Radiation} = \int d^4x \sqrt{-g}\,( {-  \sigma F^2\over 16\pi} -
W[\sigma])
   $$
where $\sigma$ is a scalar field, the density of radiation, and
$$
W[\sigma] = {1\over 3   }(\sigma\ln\sigma +1 -\sigma ).
$$
The modification of the Maxwell action by the factor $\sigma$ and the
internal energy $W$ is necessary for the same reason that we need the
factor $\rho$ and the internal energy $V$ in the matter action. It is
probable that the gravitational action should be similarly modified, to
reflect the existence of a background of soft gravitons. Compare [R] and
[AMM].

  \b\b
\no{\steptwo Acknowledgements}

I am grateful to R.J. Finkelstein and to R.W. Huff for useful
discussions.
\ve
\bb
\no {\steptwo References}
{ \settabs\+&~~~~~~~~~ &~~&\hskip5mm \cr

\+&[AMM] &Antoniadis, I., Mazur, P. and Mottola, E.,\cr 
 \+&&Cosmological Dark
Energy: Prospects for a Dynamical Theory, gr-qc/0612068.\cr

\+& [C1]&Chandrasekhar, S., The maximum mass of ideal white dwarfs,\cr
\+&&Astrophys. J. {\bf 74}, 81- (1931).\cr

\+&[C2] &Chandrasekhar, S., Stellar Configurations with Degenerate Cores,\cr
\+&& Monthly Notices R.A. {\bf 95}, 226-260 (1935).\cr

\+&[Ed] & Eddington, A.S., {\it The Internal Constitution of
Stars}, Dover, N.Y. 1959.\cr

\+&[Em]&Emden, R., {\it Gaskugeln}, Teubner 1907.\cr

\+&[F1]&Fr\o nsdal, C., Growth of a Black Hole,\cr 
\+&&J.Geom.Phys. {\bf 57}, 167-176 (2006). [gr-qc/0508048v]\cr

\+&[F2]& Fr\o nsdal,  C.,  Ideal Stars and General Relativity,
gr-qc/0606027.\cr

\+&[H] &Hartle, J.B., Bounds on mass and moment of inertia of
non-rotating\cr
\+&&stars, Physics Reports, {\bf 46}, 201-247 (1978).\cr

\+&[KH]& Kippenhahn, R. and Weigert, A, ``Stellar Structure and
Evolution",\cr
\+&&Springer-Verlag 1990.\cr

\+&[ MM]&Mazur, P. and Mottola, E., Gravitational Vacuum Condensate Stars,\cr
\+&&gr-qc/0407075\cr

\+&[OV] &Oppenheimer, J.R. and Volkoff, G.M., On Massive Neutron Stars,\cr
\+&& Phys. Rev. {\bf 55}, 347-381, (1939).\cr

\+&[R] &Rees, M.F., Effects of very long wavelength primordial gravitational radiation,\cr
\+&&Mon.Not.astr.Soc. {\bf 154} 187-195 (1971).\cr}

\ve

{\baselineskip=11.6pt

{\stephalf
$\s\nu(0)$  \hskip.8cm
   ($\s 1/a$) ~~~~~~$\s 2mG$  ~~~~~~$~~~~\s[R_{cr}]
~~~~~~~~~~~~~f(0) ~~~~2mG/a^{3/2}~~~~~~~~ ~~~ $

\settabs\+&10 .0208187~ ~~~~~   &(200)~~~~~
~~&4.94[525]~~~~~~&.025871(-)~~~~~~&1.733[144]  ~&~~~
\hskip-.3cm.00253 ~~~ (-)&~~~~.1607[13] ~&.0140868(-)
~&~~~~~~~.00185[.105]~~~&~~~~~~~~\cr

\+ &.3469775  10$^{-\s 5}$ &(825 000)~~~&1.35 10$^{-\s
9}$&[3.22 10$^{-\s 4}$]&1.07918&1.011  \cr

\+ &.4089376  10$^{-\s 5}$ &(700 000)~~~&1.72 10$^{-\s
9}$&[3.47 10$^{-\s 4}$]&1.07918&1.007 \cr

\+ &.477093666  10$^{-\s 5}$ &(600 000)~~~&2.175 10$^{-\s
9}$&[3.77 10$^{-\s 4}$]&1.07918&1.011  \cr

\+ &.5725120  10$^{-\s 5}$ &(500 000)~~~&2.83 10$^{-\s
9}$&[4.13 10$^{-\s 4}$]&1.07918&1.006 \cr

\+ &.715640106  10$^{-\s 5}$ &(400 000)~~~&4.00 10$^{-\s
9}$&[4.62 10$^{-\s 4}$]&1.07918&1.012 \cr

\+ &.95419005  10$^{-\s 5}$ &(300 000)~~~&6.155 10$^{-\s
9}$&[5.3410$^{-\s 4}$]&1.07919&1.011  \hskip1cm {\bf Table 2}\cr

\+ &.143128161  10$^{-\s 4}$ &(200 000)~~~&1.13 10$^{-\s
8}$&[6.52 10$^{-\s 4}$]&1.07919&1.011 \cr

\+ &.19083741  10$^{-\s 4}$ &(150 000)~~~&1.74 10$^{-\s
8}$&[7.58 10$^{-\s 4}$]&1.07919&1.011 \cr

\+ &.28625595  10$^{-\s 4}$ &(100 000)~~~&3.2 10$^{-\s
8}$&[9.23 10$^{-\s 4}$]&1.07919&1.012  \cr

\+ & .572512 10$^{-\s 4 }$&(50 000)~~~&.905  10$^{-\s
7}$&[.000585]&1.07921 &1.012
\cr

\+ &.00057250563&(5 000)~~~&2.9 10$^{-\s 6}$&[.00415]&1.07942 &1.025  \cr

\+ &.0023853416&(1 200)~~~&.00002425&[.00844]&1.08017&1.008  \cr

\+ &.0140868&(200)~~~&.000349&[.020]&1.08526&1.0115  \cr

\+ &.02861153&(100)&.00097&[.029]&1.09174&.97 \cr

\+ &.0476807 &(60)&.00202&[.038]&1.10101&.939 \cr

\+ &.0572184&(50)&.00261&[.042]&1.10591&.923 \cr

\+&.071534195&(40)& .00357&[.0465]&1.11359&.903 \cr

\+&  .0954368&(30)&.00529&[.0535]&1.17942&.869 \cr

\+&.1435468&(20)&.00905&[.0655]&1.1579&.809 \cr

\+& .1692845&(17)&.01110&[.0687]&1.17601&.778 \cr

\+& .18602925&(15.5)&.01244&[.0736]&1.18846&.759 \cr

\+&.1923886&(15)&.01295&[.0748]&1.19334&.752 \cr

\+& .2035445&(14.2) &.013825&[.0766]&1.20209&.740 \cr

\+ &.2230195&(13) &.01534&[.0796] &1.218&.719 \cr

\+ &.242466&(12) &.01680&[.0825]&1.23475&.698 \cr

\+ &.2658189&(11) &.01853&[.0855]&1.25606&.676 \cr

\+ &.2944944&(10) &.02053&[.0884]&1.28418&.649 \cr

\+ &.330789&(9) &.02286&[.0920]&1.3232&.617 \cr

\+ &.3788565&(8) &.02550&[.0950]&1.3818&.577 \cr

\+ &.447908&(7) &.02846&[.0980]&1.4832&.527 \cr

\+ &.575115&(6) &.03060&[.0962]&1.749&.450 \cr

\+ &.67760&(5.75857) &.0290&[.0910]&2.09284&.401 \cr

\+ &.7529185&(5.9) &.02596&[.0840]&2.48505&.372 \cr

\+ &.790723&(6.1) &.02390&[.0793]&2.7564&.360 \cr

\+ &.874&(7) &.01833&[.0680]&3.6683&.339 \cr

\+ &.92404&(8) &.01466&[.0605]&4.5583&.332 \cr

\+ &.98369&(10) &.01039&[.0512]&6.2787&.329 \cr

\+ &1.00924&(11.245) &.00874&[.0474]&7.3522&.330 \cr

\+ &1.013842&(11.5) &.00846&[.0468]&7.5735&.330 \cr

\+ &1.022385&(12) &.00795&[.0455]&8.0092&.330 \cr

\+ &1.051813&(14) &.0064&[.042 ]&9.780&.335 \cr

\+ &1.119438&(20) &.00395&[.0353]&14.473&.353 \cr

\+ &1.17798&(25) &.00305&[.0332]&21.073&.381 \cr

\+ &1.19458&(26) &.00295&[.0331]&22.446&.391 \cr

\+ &1.2183451&(27) &.0028&[.0332]&24.114&.393 \cr}}

\b
  {\stephalf
\ce{\bf Table 3, $n=3,15$}
\vskip.1cm
\settabs\+&10 .0208187~ ~~ ~~ 看 &(200)~~~ ~~~~
~&4.94[525]~~~~~~~~&.025871(-)~~~~~~&1.733[144] 看&
\hskip-.3cm.00253 看(-)&~.1607[13] ~&.0140868(-)
~&~~~~.00185[.105]&~~~~~\cr

\+&~~~$ \s\nu(0)$ 看\hskip1.2cm
  看($\s 1/a$) ~~~~~~~~~~$\s 2mG$ 看~~~~~~$~~~~[\s R_{cr}]
~~~~~~~~~~~~~ f_{cr} ~~~~~~~~~~~ $\cr

\+ &5.821298 10$^{-6}$ &(540 000)~~~&2.32 10$^{-9}$
&[5.5 10$^{-4}$]&1.644&  \cr

\+ &3.146561 10$^{-4}$ &(10 000)~~~&9.2 10$^{-7}$
&[.004] &1.64887 & \cr

\+&.0314001 &(100) ~~~&8.82 10$^{-4}$ &[.04]
&1.712 &  \cr

\+&.06271548 &(50) & .002387 & [.053] & 1.7876
&  \cr

\+&.15696 & (20) & .00828 & [.08] & 2.0708
&   \cr

\+&.3221575 & (10) & .01878 & [.1] & 2.896
&  \cr

\+&.4165655 & (8) & 0.2326 & [.103] & 3.7255
&   \cr

\+&.4980215 & (7) & .02572 & [.104] & 4.864
&  \cr

\+&.64313 & (6.2) & .0262 & [.1] & 9.236
& \cr

\+&.787609 & (6.5) & .02135 & [.08] & 24.91
&  \cr

\+&.965793 & (10) & .01026 & [.055] & 227.0
& \cr

\+&1.1022975 & (15) & .00386 & [.038] & 3431.4
&  \cr

\+&1.150818 & (25) & .00292 & [.035] & 8312.7
& \cr}

\b

{\stephalf
\ce{\bf Table 4, $n=2.5,6$}
\vskip.1cm
\settabs\+&10 .0208187~  ~~~~~~ 看 &(200)~~~~~~~
~&4.94[525]~~~~~~~~~&.025871(-)~~~~~&1.733[144] 看&
\hskip-.3cm.00253 看(-)&~.1607[13] ~&.0140868(-)
~&~~~~.00185[.105]&~~~~~\cr

\+&~~~$\s \nu(0)$ 看\hskip1.2cm
  看~($\s 1/a$) ~~~~~~~~$\s  2mG$ 看~~~~~~$~~~~~~[\s R_{cr}]
~~~~~~~~~~~~~ f_{cr} ~~~~~~~~~~~ $\cr

\+&0.5582799 & (50.5) & .0026 & [.0401] & 1.0578
&   \cr

\+&.0560465 & (50.3) & .00261 & [.0401] & 1.0578
&   \cr

\+&.056381205 & (50) & .002642 & [.0403] & 1.0579
&  \cr

\+& &(35) & .00435 & [.0481] & 1.0637 &  \cr

\+&.1399534 & (20) & .00921 & [.064] & 1.0813
&  \cr

\+&.231989 & (12) & .01735 & [.0815] & 1.1205
&   \cr

\+&.309115 & (9) & .02400 & [.0924] & 1.1645
&   \cr

\+& .3995988 & (7) & .0310 & [.1012] & 1.2281
&  \cr

\+&.4708388 & (6) & .0357 & [.1068] & 1.2886
&  \cr

\+&.580451 & (5) & .0411 & [.111] & 1.4051
&  \cr

\+&.835304 & (4) & .0442 & [.108] & 1.865
&   \cr

\+&& (3.95) & .04318 & [.1053] & 8.152 &   \cr

\+&.9446 & (3.95) & .0414 & [.1016] & 2.2173 &   \cr

\+&.99335 & (4) & .03946 & [.092] & 2.429 &  \cr

\+&1.213762 & (5) & .0267 & [.0775] & 4.2252 &   \cr
\+&1.309025 & (6) & .02094 & [.069] & 5.7912 &  \cr
\+&1.380006 & (7) & .01735 & [.064] & 7.437 & \cr
\+&1.412063 & (7.5) & .01601 & [.0621] & 8.315 &  \cr
\+&1.449928  & (8.1) & .01472 & [.0605] & 9.439 &  \cr
\+&1.510485 & (9) & .01424 & [.0595] & 11.34 &   \cr
\+&1.55061 & (9.5) & .012625 & [.059] & 12.60 &   \cr
\+&1.60489 & (10) & .01218 & [.0586] & 14.20 &   \cr
\+&1.62 & (10.1) & .01213 & [.0588] & 14.62 &  \cr}

\ve

{\stephalf

\vskip.1cm
  \settabs\+&10 .0208187~ ~~~~~~~ 看 &(200)~~~~~~~
~&4.94[525]~~~~~~~~~&.025871(-)~~~~~&1.733[144] 看&
\hskip-.3cm.00253 看(-)&~.1607[13] ~&.0140868(-)
~&~~~~.00185[.105]&~~~~~\cr

\+&~~~$\s \nu(0)$ 看\hskip1.2cm
  看($\s 1/a$) ~~~~~~~~~~~$\s 2mG$ 看~~~~~~~$~~~~[\s R_{cr}]
~~~~~~~~~~~~~ f_{cr} ~~~~~~~~~~~ $\cr

\+ && (~) & .02127 & [.085] & 1.06765 &   \cr
\+&.272783 & (10) & .0217 & [.0866] & 1.06864 &  \cr
\+&.286664 & (9.5) & .023 & [.0886] & 1.07172 &   \cr
\+&.3020432 & (9) & .02441 & [.0908] & 1.0754 &  \cr
\+&.3384081 & (8) & .0277 & [.0955] & 1.08537 &   {\bf Table 5, $n=2,6$} \cr
\+&.3601397 & (7.5) & .02967 & [.0982] & 1.09232 & \cr
\+&.3849225 & (7) & .03184 & [.1008] & 1.10127 &  \cr
\+&.4134867 & (6.5) & .03425 & [.1037] & 1.11298 &  \cr
\+&.4468366 & (6) & .037 & [.107] & 1.12849 &   \cr
\+&.5343378 & (5) & .04356 & [.1135] & 1.17742 &  \cr
\+&.671471 & (4) & .0518 & [.120] & 1.2762 &   \cr
\+&.9584545 & (3) & .05972 & [.122] & 1.6077 &  \cr
\+&1.15677 & (2.78) & .0579 & [.1154] & 2.0198 &   \cr
\+&1.24684 & (2.78) & .0552 & [.111] & 2.2975 &  \cr
\+&1.434277 & (3) & .0471 & [.0995] & 3.1966 &  \cr
\+&1.7143226 & (4) & .0333 & [.0833] & 6.071 &  \cr
\+&1.91458 & (5) & .02714 & [.0781] & 9.6402 &   \cr
\+&2.03779 & (5.45) & .02569 & [.078] & 12.124 &  \cr

\b
\hrule
\vskip.1cm
  \settabs\+&10 .0208187~ ~~~~~~ 看 &(200)~~~~~~~
~&4.94[525]~~~~~~~~~&.025871(-)~~~~~~&1.733[144] 看&
\hskip-.3cm.00253 看(-)&~.1607[13] ~&.0140868(-)
~&~~~~.00185[.105]&~~~~~\cr

\+&~~~$\s \nu(0)$ 看\hskip.9cm
  看~~~~($\s 1/a$) ~~~~~~~~~$\s 2mG$ 看~~~~~~$~~~~~~[\s R_{cr}]
~~~~~~~~~~~~~ f_{cr} ~~~~~~~~~~~ $\cr

\+&& (50) & .00266 & [.0367] & 1.02525 &   \cr
\+&.0925715 & (30) & .00542 & [.0481] & 1.0279 &   \cr
\+&.11082561 & (25) & .00696 & [.053] & 1.02922 &   \cr
\+&.18293098 & (15) & .01347 & [.069] & 1.03453 &  \cr
\+&.2272555 & (12) & .0177 & [.0772] & 1.03793 &  \cr
\+&.2472169 & (11) & .0196 & [.08055] & 1.03951 &   \cr
\+&.271095 & (10) & .0218 & [.0843] & 1.04145 &   \cr
\+&.2998917 & (9) & .0246 & [.0885] & 1.0439 &  {\bf Table 6, $n=1.5,6$} \cr
\+&.335651 & (8) & .0279 & [.0933] & 1.0471 &  \cr
\+&.3811072 & (7) & .0321 & [.099] & 1.05151 &   \cr
\+&.4088025 & (6.5) & .0346 & [.102] & 1.05444 &  \cr
\+&.4408607 & (6) & .0374 & [.105] & 1.05813 &  \cr
\+&.478416 & (5.5) & .0406 & [.1085] & 1.06296 &   \cr
\+&.5230489 & (5) & .04425 & [.1124] & 1.06966 &   \cr
\+&.5770453 & (4.5) & .0484 & [.1164] & 1.07976 &   \cr
\+&.643911 & (4) & .0532 & [.1205] & 1.09704 &   \cr
\+&.729582 & (3.5) & .0587 & [.1245] & 1.1308 &  \cr
\+&.8457205 & (3) & .0649 & [.1285] & 1.19687 &   \cr
\+&1.02041 & (2.5) & .0712 & [.1308] & 1.33075 &   \cr
\+&1.40847 & (2) & .0735 & [.1251] & 1.85382 &   \cr
\+&1.715 & (2) & .0658 & [.1132] & 2.734 &   \cr
\+&2.112565 & (2.5) & .05086 & [.0977] & 5.452 &   \cr
\+&2.358166 & (3) & .0441 & [.0928] & 8.6144 &   \cr
\+&2.467983 & (3.2) & .0424 & [.0923] & 10.367 &   \cr}

\ve

{\stephalf
\ce{\bf Table 7, $n=1,6$}
\vskip.5cm
  \settabs\+&10 .0208187~ ~~~~ 看 &(200)~~~
~~~~~&4.94[525]~~~~~~~~~&.025871(-)~~~~~~&1.733[144] 看&
\hskip-.3cm.00253 看(-)&~.1607[13] ~&.0140868(-)
~&~~~~.00185[.105]&~~~~~\cr

\+&~~~$\s \nu(0)$ 看\hskip1. cm
  看($\s 1/a$) ~~~~~~~~$\s 2mG$ 看~~~~~~$~~~~[\s R_{cr}]
~~~~~~~~~~~~~~ f_{cr}  ~~~~~~~~~~~ $\cr

\+&.02791529 & (100) & .000991 & [.0237] & 1.01622 &  \cr
\+&.11050725 & (25) & .00698 & [.0501] & 1.02071 &   \cr
\+&& (20) & .00938 & [.0566] & 1.02212 &   \cr
\+&.1823652 & (15) & .01353 & [.066] & 1.02441 &  \cr
\+&.2700904 & (10) & .02197 & [.0814] & 1.02883 &   \cr
\+&.2988242 & (9) & .0247 & [.0852] & 1.03027 &   \cr
\+&.4388956 & (6) & .03766 & [.1026] & 1.03747 &   \cr
\+&.5202716 & (5) & .04457 & [.1101] & 1.04193 &   \cr
\+&.639032 & (4) & .0537 & [.1187] & 1.04909 &   \cr
\+&.721724 & (3.5) & .05936 & [.1232] & 1.05483 & \cr
\+&.8296405 & (3) & .0658 & [.1278] & 1.0639 &  \cr
\+&.9772556 & (2.5) & .0733 & [.132] & 1.08228 &  \cr
\+&1.1965135 & (2) & .08135 & [.1346] & 1.1563 &   \cr
\+&1.630761 & (1.5) & .0869 & [.1314] & 1.54041 &   \cr
\+&1.842675 & (1.4) & .0854 & [.1268] & 1.85969 &  \cr
\+&2.250955 & (1.4) & .0773 & [.1157] & 2.79388 &  \cr
\+&2.47326 & (1.5) & .07144 & [.1098] & 4.07289 &  \cr
\+&2.854157 & (1.8) & .0625 & [.1032] & 7.36191 &   \cr
\+&2.966246001 & (1.9) & .0607 & [.1022] & 8.74897 &   \cr
\+&3.0853946 & (2) & .0592 & [.102] & 10.438 & \cr}

\end

\b
\ve  \b\ce{\bf Table 4. Critical mass/radius,  new equation of
state, $n = (1,6)$ }
   \def\s{\scriptstyle}{\stephalf
~~
\settabs \+ ~     & 1~~~~~    ~~~~& 20~~~~~ ~ ~& 100~~~~~~~~&
1000~~~~~~~& 10000~~~&~~~~~~~~~ &
100000~~~~&~~~~~~~~~~~~~&~~~~\~~~~~~~&~~~~~~~~~~~&~~~~~~~~~~~&~~~~~~~~~~~~~&~~~~~~~~~\cr

\+&$-\s\nu(0)$&.003282&.006356&.012593&.03099&.06101&.0758083&.1002139&.119538
&.148233&.168537&.173295&.1953847  \cr
\+ &$\s R$&5.1&5.05&5.0&4.8&4.6&4.5&4.35&4.25&4.1& 4.01&4 &3.9 \cr
   \+&$\s2m$&.0062&.0127&.0240&.062 &.115&.139&.180&.208&.25&.276&.28&
.311\cr
\+&$\s
f_{cr}$&.00051&.001&.002&.005&.001&.0125&.0167&.02&.025&.029&.029&.033&
\cr
\+&f(0)&.00082&.00159&.00317&.00787&.0157&.0197&.0264&.0317&.0399&
.0459&.0473&.0539\cr
\b
\+&$-\s\nu(0)$&
.232&.2571607&.2876505
&.377815 &.448811 &.553973 &.727573 &1.084&1.21301  &1.3933 &
1.725148&1.9347634&
   \cr
\+ &$\s R$&3.8& 3.7&3.6&3.3&3.1&2.8&2.4&1.9&1.8&1.6&1.35&1.3 \cr
   \+&$\s2m$ &.351&.377&.407 &.48&.525&.575&62&.64&.63&.615&.56&.525\cr
\+&$\s f_{cr}$&.04&.044&.05&.067&.08&.10&.133&.2&.222&.25&.29&.29&\cr
\+&
f(0)&.065&.0733&.083&.115&.142&.185&.268&.489&.591&.757&1.15&1.48\cr
}

\b
The mass relation is plotted in Fig.2.

Typical behaviour of some of the functions is shown  in Fig.s 3 and 4.
\b
\b

\epsfxsize.7\hsize
\centerline{\epsfbox{Fig.13.eps}}
\
\vskip-1cm
\vskip-1in
\no{\it
Fig.3. On the left, typical behaviour of the metric functions $-\nu$
(upper
curve) and
$\lambda$ (lower curve). From the point where they join, the metric is
Schwartzschild to a very good approximation. The new equation of state
was
used. On the right is shown the change over of the index from 3 to 6.
    The other curve in this figure is $100f(r)$. The
values of the parameters for this particular case are
$R = 2.8, ~f_{cr} = f_{cr} = .1 $ and $f(0) = .185$.
.}
\epsfxsize.7\hsize
\centerline{\epsfbox{Fig.11.eps}}
\
\vskip-1cm
\vskip1cm
\no{\it
Fig 4. The same as Fig.3 except that the index is 10 in the atmosphere.
The
upper curve shows $1000 f(r)$. The parameters are $R = 3.4, ~f_{cr} =
1/30$
and $f(0) = 1$..}

   \b
   \no

Oscillations around the equilibrium configurations affect the metric as
well as density and pressure. The answer to the question about the mass
is
that it does not take part in these oscillations. It is a true constant
of
the motion.
\b
\ce{\bf What next?}

The factor $a$ in the equation of state was treated as a fixed
parameter and
this prevents us from applying the theory to actual stellar objects. An
examination of the values of the density at the center can help to
fix the scale and to overcome this difficulty. Compare [H].\break

\b\b
\no{\steptwo Acknowledgements}

I am grateful to R.J. Finkelstein and to R.W. Huff for useful
discussions.

\bb
\no {\steptwo References}

[C1] ~Chandrasekhar, S., The maximum mass of ideal white dwarfs,

\quad\quad~ Astrophys. J. {\bf 74}, 81- (1931).

[C2] ~Chandrasekhar, S., Stellar Configurations with Degenerate Cores,

\quad\quad  \hskip2mm Monthly
Notices R.A. {\bf 95}, 226-260 (1935).

{Ed]~\hskip2.2mm  Eddington, A.S., \it The Internal Constitution of
Stars},
Dover, N.Y. 1959.

[Em]\hskip.2mm ~Emden, R., {\it Gaskugeln}, Teubner 1907.

[F] ~~\hskip.8mmFr\o nsdal,  C.,  Ideal Stars and General Relativity,
gr-qc/0606027.

[H] \hskip2.8mmHartle, J.B., Bounds on mass and moment of inertia of
non-rotating

\quad\quad\hskip2mmneutron stars, Physics Reports, {\bf 46}, 201-247
(1978).

[KH]\hskip1.3mm  Kippenhahn, R. and Weigert, A, ``Stellar Structure and
Evolution",

   \quad \quad\hskip1mm ~Springer-Verlag 1990.

[OV] Oppenheimer, J.R. and Volkoff, G.M., On Massive Neutron Stars,

\quad\quad \hskip2mm  Phys.
Rev. {\bf 55}, 347-381, (1939).

\ve

   \vskip1.1in
\def\picture #1 by #2 (#3){
   \vbox to #2{
     \hrule width #1 height 0pt depth 0pt
     \vfill
     \special{picture #3} 
     }
   }
\def\scaledpicture #1 by #2 (#3 scaled #4){{
   \dimen0=#1 \dimen1=#2
   \divide\dimen0 by 1000 \multiply\dimen0 by #4
   \divide\dimen1 by 1000 \multiply\dimen1 by #4
   \picture \dimen0 by \dimen1 (#3 scaled #4)}
   }

\parindent=1pc

\epsfxsize.7\hsize
\centerline{\epsfbox{Fig.12.eps}}

\vskip-1cm
\vskip0cm
\no{\it Fig.1.  The mass/radius relation . Here the polytropic index in the
bulk is
3. The upper curve is the result obtained by joining the interior
solution
to an exterior Schwartzschild metric. The other curve was obtained by
using
the new equation of state with index 3 (interior) and 6 (exterior).
The abcissa is the natural logarithm $\ln(R)$ of the radius $R$ and the
ordinate is $2m$.}

\parindent=1pc

\epsfxsize.7\hsize
\centerline{\epsfbox{Fig.10.eps}}

\vskip1cm
\vskip1cm
\no{\it
Fig.2. The critical mass/radius relation for the case that the
polytropic
index is 1 in the bulk. The abcissa is
the radius $R$ (not the logarithm) and the ordinate is $2m$. The ratio
$2m/R$
in cgs units is the dimensionless number $2mG/R$. On the left the
result
obtained previously by joining a polytrope with index 1 to an exterior
Schwartzschild metric. On the right the relation obtained using the new
equation of state with a polytropic index that changes from 1 (inside)
to 6
at the point of critical density.}

\epsfxsize.7\hsize
\centerline{\epsfbox{Fig.13.eps}}
\
\vskip-1cm
\vskip1cm
\no{\it
Fig.3. On the left, typical behaviour of the metric functions $-\nu$
(upper
curve) and
$\lambda$ (lower curve). From the point where they join, the metric is
Schwartzschild to a very good approximation. The new equation of state
was
used. On the right is shown the change over of the index from 3 to 6.
    The other curve in this figure is $100f(r)$. The
values of the parameters for this particular case are
$R = 2.8, ~f_{cr} = f_{cr} = .1 $ and $f(0) = .185$.
.}
\epsfxsize.7\hsize
\centerline{\epsfbox{Fig.11.eps}}
\
\vskip-1cm
\vskip1cm
\no{\it
Fig 4. The same as Fig.3 except that the index is 10 in the atmosphere.
The
upper curve shows $1000 f(r)$. The parameters are $R = 3.4, ~f_{cr} =
1/30$
and $f(0) = 1$..}

\end
\no{\bf Figure captions.}

Fig.1.  The mass/radius relation . Here the polytropic index in the
bulk is
3. The upper curve is the result obtained by joining the interior
solution
to an exterior Schwartzschild metric. The other curve was obtained by
using
the new equation of state with index 3 (interior) and 6 (exterior).
The abcissa is the natural logarithm $\ln(R)$ of the radius $R$ and the
ordinate is $2m$.

Fig.2. The critical mass/radius relation for the case that the
polytropic
index is 1 in the bulk. The abcissa is
the radius $R$ (not the logarithm) and the ordinate is $2m$. The ratio
$2m/R$
in cgs units is the dimensionless number $2mG/R$. On the left the
result
obtained previously by joining a polytrope with index 1 to an exterior
Schwartzschild metric. On the right the relation obtained using the new
equation of state with a polytropic index that changes from 1 (inside)
to 6
at the point of critical density.

\b
Fig.3. On the left, typical behaviour of the metric functions $-\nu$
(upper
curve) and
$\lambda$ (lower curve). From the point where they join, the metric is
Schwartzschild to a very good approximation. The new equation of state
was
used. On the right is shown the change over of the index from 3 to 6.
    The other curve in this figure is $100f(r)$. The
values of the parameters for this particular case are
$R = 2.8, ~f_{cr} = f_{cr} = .1 $ and $f(0) = .185$.

\b

Fig 4. The same as Fig.3 except that the index is 10 in the atmosphere.
The
upper curve shows $1000 f(r)$. The parameters are $R = 3.4, ~f_{cr} =
1/30$
and $f(0) = 1$.

  \b\b
\no{\steptwo Acknowledgements}

I am grateful to R.J. Finkelstein and to R.W. Huff for useful 
discussions.

\bb
\no {\steptwo References}

[C1] ~Chandrasekhar, S., The maximum mass of ideal white dwarfs,

\quad\quad~ Astrophys. J. {\bf 74}, 81- (1931).

[C2] ~Chandrasekhar, S., Stellar Configurations with Degenerate Cores,

\quad\quad  \hskip2mm Monthly
Notices R.A. {\bf 95}, 226-260 (1935).

{Ed]~\hskip2.2mm  Eddington, A.S., \it The Intenal Constitution of 
Stars},
Dover, N.Y. 1959.

[Em]\hskip.2mm ~Emden, R., {\it Gaskugeln}, Teubner 1907.

[F] ~~\hskip.8mmFr\o nsdal,  C.,  Ideal Stars and General Relativity,
gr-qc/0606027.

[H] \hskip2.8mmHartle, J.B., Bounds on mass and moment of inertia of
non-rotating

\quad\quad\hskip2mmneutron stars, Physics Reports, {\bf 46}, 201-247
(1978).

[KH]\hskip1.3mm  Kippenhahn, R. and Weigert, A, ``Stellar Structure and
Evolution",

  \quad \quad\hskip1mm ~Springer-Verlag 1990.

[OV] Oppenheimer, J.R. and Volkoff, G.M., On Massive Neutron Stars,

\quad\quad \hskip2mm  Phys.
Rev. {\bf 55}, 347-381, (1939).

\ve
\end

  \vskip1.1in
\def\picture #1 by #2 (#3){
  \vbox to #2{
    \hrule width #1 height 0pt depth 0pt
    \vfill
    \special{picture #3} 
    }
  }
\def\scaledpicture #1 by #2 (#3 scaled #4){{
  \dimen0=#1 \dimen1=#2
  \divide\dimen0 by 1000 \multiply\dimen0 by #4
  \divide\dimen1 by 1000 \multiply\dimen1 by #4
  \picture \dimen0 by \dimen1 (#3 scaled #4)}
  }

\parindent=1pc

\epsfxsize.7\hsize
\centerline{\epsfbox{Fig.12.eps}}
 
\vskip-1cm
\vskip0cm
\no{\it Fig.1.  The mass/radius relation . Here the polytropic index in the 
bulk is
3. The upper curve is the result obtained by joining the interior 
solution
to an exterior Schwartzschild metric. The other curve was obtained by 
using
the new equation of state with index 3 (interior) and 6 (exterior).
The abcissa is the natural logarithm $\ln(R)$ of the radius $R$ and the
ordinate is $2m$.}

\parindent=1pc

\epsfxsize.7\hsize
\centerline{\epsfbox{Fig.10.eps}}
 
\vskip1cm
\vskip1cm
\no{\it 
Fig.2. The critical mass/radius relation for the case that the 
polytropic
index is 1 in the bulk. The abcissa is
the radius $R$ (not the logarithm) and the ordinate is $2m$. The ratio 
$2m/R$
in cgs units is the dimensionless number $2mG/R$. On the left the 
result
obtained previously by joining a polytrope with index 1 to an exterior
Schwartzschild metric. On the right the relation obtained using the new
equation of state with a polytropic index that changes from 1 (inside) 
to 6
at the point of critical density.}

\epsfxsize.7\hsize
\centerline{\epsfbox{Fig.13.eps}}
\
\vskip-1cm
\vskip1cm
\no{\it 
Fig.3. On the left, typical behaviour of the metric functions $-\nu$ 
(upper
curve) and
$\lambda$ (lower curve). From the point where they join, the metric is
Schwartzschild to a very good approximation. The new equation of state 
was
used. On the right is shown the change over of the index from 3 to 6.
   The other curve in this figure is $100f(r)$. The
values of the parameters for this particular case are
$R = 2.8, ~\tau = f_c = .1 $ and $f(0) = .185$.
.}    
\epsfxsize.7\hsize
\centerline{\epsfbox{Fig.11.eps}}
\
\vskip-1cm
\vskip1cm
\no{\it 
Fig 4. The same as Fig.3 except that the index is 10 in the atmosphere. 
The
upper curve shows $1000 f(r)$. The parameters are $R = 3.4, ~\tau = 
1/30$
and $f(0) = 1$..}

\end
\no{\bf Figure captions.}

Fig.1.  The mass/radius relation . Here the polytropic index in the 
bulk is
3. The upper curve is the result obtained by joining the interior 
solution
to an exterior Schwartzschild metric. The other curve was obtained by 
using
the new equation of state with index 3 (interior) and 6 (exterior).
The abcissa is the natural logarithm $\ln(R)$ of the radius $R$ and the
ordinate is $2m$.

Fig.2. The critical mass/radius relation for the case that the 
polytropic
index is 1 in the bulk. The abcissa is
the radius $R$ (not the logarithm) and the ordinate is $2m$. The ratio 
$2m/R$
in cgs units is the dimensionless number $2mG/R$. On the left the 
result
obtained previously by joining a polytrope with index 1 to an exterior
Schwartzschild metric. On the right the relation obtained using the new
equation of state with a polytropic index that changes from 1 (inside) 
to 6
at the point of critical density.

\b
Fig.3. On the left, typical behaviour of the metric functions $-\nu$ 
(upper
curve) and
$\lambda$ (lower curve). From the point where they join, the metric is
Schwartzschild to a very good approximation. The new equation of state 
was
used. On the right is shown the change over of the index from 3 to 6.
   The other curve in this figure is $100f(r)$. The
values of the parameters for this particular case are
$R = 2.8, ~\tau = f_c = .1 $ and $f(0) = .185$.

\b

Fig 4. The same as Fig.3 except that the index is 10 in the atmosphere. 
The
upper curve shows $1000 f(r)$. The parameters are $R = 3.4, ~\tau = 
1/30$
and $f(0) = 1$.

\def\picture #1 by #2 (#3){
   \vbox to #2{
     \hrule width #1 height 0pt depth 0pt
     \vfill
     \special{picture #3} 
     }
   }

\def\scaledpicture #1 by #2 (#3 scaled #4){{
   \dimen0=#1 \dimen1=#2
   \divide\dimen0 by 1000  \multiply\dimen0 by #4
   \divide\dimen1 by 1000 \multiply\dimen1 by #4
   \picture \dimen0 by \dimen1 (#3 scaled #4)}
   }